\pdfoutput=1 % only if pdf/png/jpg images are used
\documentclass{JINST}

\title{Improvement of Spectroscopic Performance \newline using a Charge-sensitive Amplifier Circuit \newline for an X-Ray Astronomical SOI Pixel Detector}

\author{Ayaki~Takeda$^a$\thanks{Corresponding author.}~, Takeshi~Go~Tsuru$^a$, Takaaki~Tanaka$^a$, Hiroyuki~Uchida$^a$, Hideaki~Matsumura$^a$, Yasuo~Arai$^b$, Koji~Mori$^c$, Yusuke~Nishioka$^c$, Ryota~Takenaka$^c$, Takayoshi~Kohmura$^d$, Shinya~Nakashima$^e$, Shoji~Kawahito$^f$, Keiichiro~Kagawa$^f$, Keita~Yasutomi$^f$, Hiroki~Kamehama$^f$, and Sumeet~Shrestha$^f$\\
\llap{$^a$}Department of Physics, Faculty of Science, Kyoto University,\\
  Kitashirakawa-Oiwakecho, Sakyo-ku, Kyoto 606-8502, Japan\\
\llap{$^b$}High Energy Accelerator Research Organization (KEK),\\
  1-1 Oho, Tskuba, Ibaraki 305-0801, Japan\\
\llap{$^c$}Department of Applied Physics, Faculty of Engineering, University of Miyazaki,\\
  1-1 Gakuen-Kibanodai-Nishi, Miyazaki, Miyazaki 889-2192, Japan\\
\llap{$^d$}Department of Physics, School of Science and Technology, Tokyo University of Science,\\
  2641 Yamazaki, Noda, Chiba 278-8510, Japan\\
\llap{$^e$}Japan Aerospace Exploration Agency (JAXA),\\
  3-1-1 Yoshinodai, Sagamihara, Kanagawa 229-8510, Japan\\
\llap{$^f$}Research Institute of Electronics, Shizuoka University,\\
  3-5-1, Johoku, Naka-ku, Hamamatsu, Shizuoka 432-8011, Japan\\
E-mail: \email{atakeda@cr.scphys.kyoto-u.ac.jp}}

\abstract{
We have been developing monolithic active pixel sensors series, named ``XRPIX,''~based on the silicon-on-insulator (SOI) pixel technology, for future X-ray astronomical satellites. 
The XRPIX series offers high coincidence time resolution (${\rm \sim}$1~${\rm \mu s}$), superior readout time (${\rm \sim}$10~${\rm \mu s}$), and a wide energy range (0.5--40 keV). 
In the previous study, we successfully demonstrated X-ray detection by event-driven readout of XRPIX2b.
We here report recent improvements in spectroscopic performance.
We successfully increased the gain and reduced the readout noise in XRPIX2b by decreasing the parasitic capacitance of the sense-node originated in the buried p-well (BPW).
On the other hand, we found significant tail structures in the spectral response due to the loss of the charge collection efficiency when a small BPW is employed. 
Thus, we increased the gain in XRPIX3b by introducing in-pixel charge sensitive amplifiers instead of having even smaller BPW.  
We finally achieved the readout noise of 35 ${e^{-}}$ (rms) and the energy resolution of 320 eV (FWHM) at 6 keV without significant loss of the charge collection efficiency. 
}

\keywords{silicon-on-insulator (SOI); SOIPIX; charge sensitive amplifier (CSA) pixel circuit; spectroscopic performance; X-ray astronomy}

\begin{document}

\section{Introduction}
\label{sec:introduction}

Charge-coupled devices (CCDs) \cite{chandra_ccd}-\cite{koyama_pasj07} are the standard imaging spectrometers in current X-ray astronomical satellites.
They offer Fano-limited spectroscopic performance with a low readout noise of about three electrons (rms) and superior imaging performance with the sensor size of 20--30 mm and a pixel size of ${\rm \sim}$24~${\rm \mu m}$.
However, X-ray CCDs have some limitations.
The most significant issue is the high non-X-ray background (NXB) above 10~keV generated by the cosmic rays in orbit. 

\begin{figure}[!tb]
	\centering
	\vspace*{-2\intextsep}
	\includegraphics[width=12.5cm]{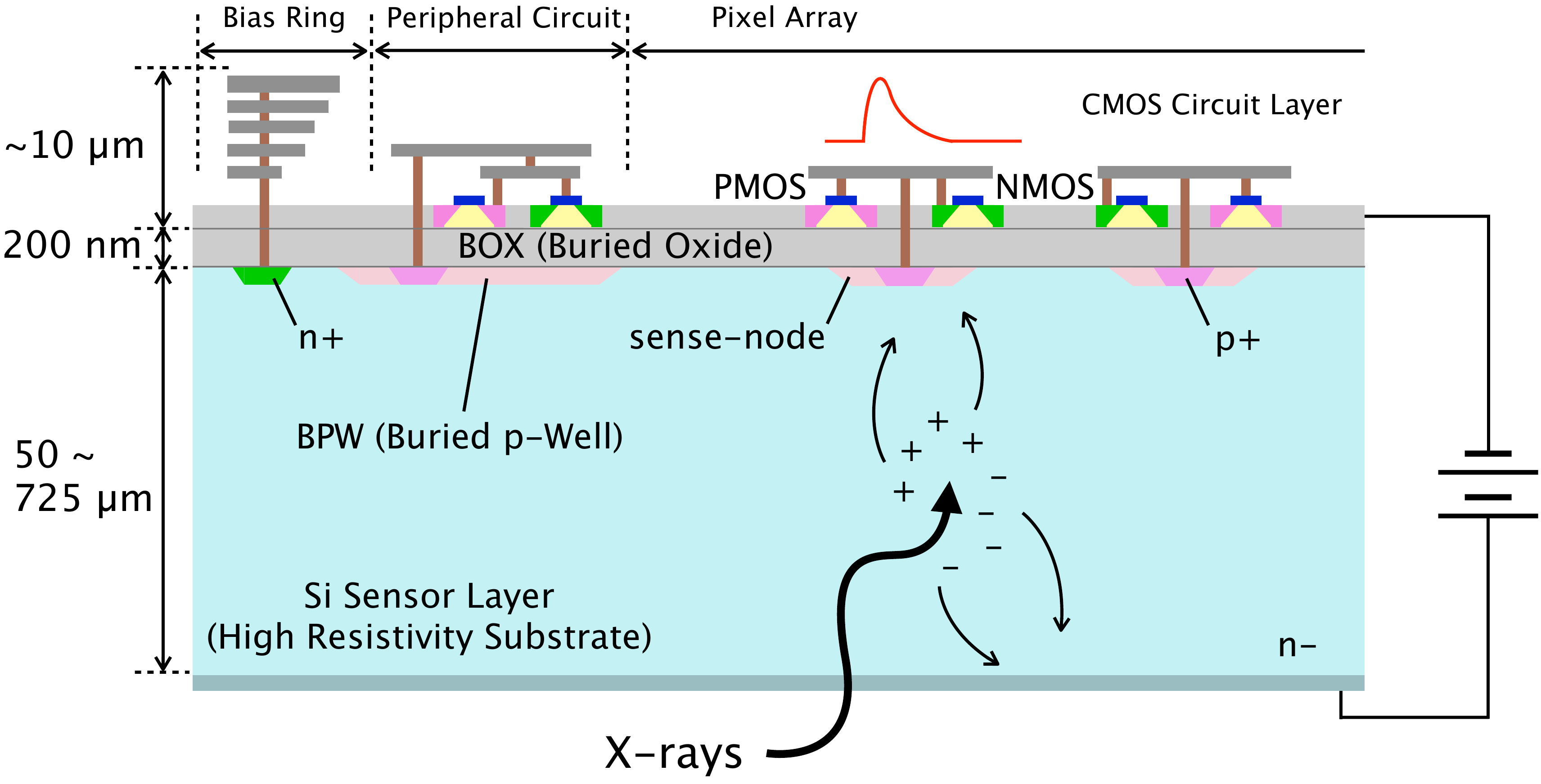}
	\caption{The cross-sectional view of SOIPIX.}
	\label{fig:soipix}
	\vspace*{-\intextsep}
\end{figure}

The NXB can be significantly reduced by introducing an anti-coincidence method between hit signals and external active shield detectors \cite{takahashi_07}-\cite{anada_08}.
This technique requires an X-ray sensor with a high coincidence time resolution (i.e.,~${\rm \sim}$1 ${\rm \mu}$s) and short readout time (i.e.,~${\rm \sim}$10~${\rm \mu}$s).
Thus, we started developing a new type of active pixel sensors, named XRPIX (X-Ray PIXel). 
The XRPIX series is one of the ``SOIPIX'' family, which is a active pixel sensor based on a semiconductor pixel detector realized in silicon-on-insulator (SOI) complementary metal-oxide-semiconductor (CMOS) pixel technology led by the High Energy Accelerator Research Organization (KEK)~\cite{arai_nima11}. 
Figure~\ref{fig:soipix} shows the cross-sectional view of SOIPIX.
The SOI wafer composed of two bonded silicon wafers with a thin oxide film (BOX: buried oxide) in between.
The pixel detector consists of a circuit built within the low-resistivity silicon and a sensing volume realized in the high-resistivity silicon.
The SOIPIX utilizes the thick handle wafer of the SOI structure as a sensing volume to detect X-rays.
We have processed four prototype productions of XRPIX1, XRPIX1b, XRPIX2 and XRPIX2b, and reported the results with them \cite{ryu_tns11}-\cite{matsumura_nima15}. 
In our previous study, we successfully demonstrated the X-ray detection by the event-driven readout \cite{takeda_tns13}\cite{takeda_pos14}.
Here, we report recent improvements in spectroscopic performance in this paper.

\enlargethispage{5mm}

The three major factors determining the spectral performance are readout noise, charge collection, and dark current.
The noise due to the dark current is simply suppressed by cooling. 
The charge collection related issues of XRPIX are studied in \cite{matsumura_nima15}\cite{matsumura_nima14}.
In this paper, we focus on reduction of readout noise mainly limited by in-pixel and on-chip readout circuits. 
We investigate a relation between gain and readout noise.
Then, we describe a new prototype, ``XRPIX3b,'' which has a charge-sensitive amplifier (CSA) circuit in each pixel in order to increase the conversion gain and reduce readout noise.
We also provide some details of the next design challenge from the information obtained.

\section{Reduction of Readout Noise by Small BPW}
\label{sec:spectroscopic_performance}

\begin{figure}[!tb]
	\centering
	\vspace*{\intextsep}
	\includegraphics[width=15cm]{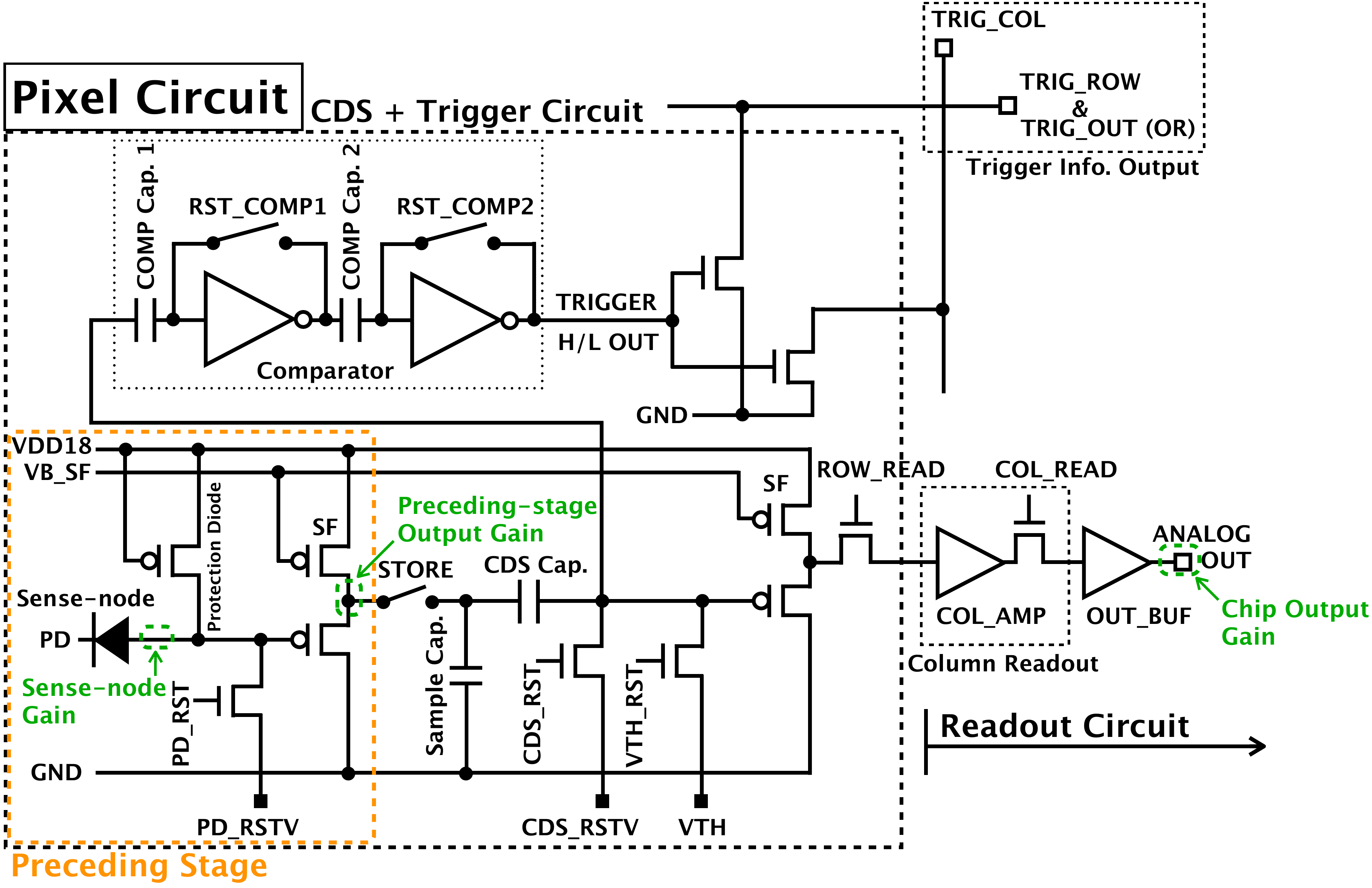}
	\caption{Pixel circuit of XRPIX2b. It has a CDS circuit like the former device. The comparator for trigger detection is inverter chopper type with preset threshold voltage. A location of ``preceding stage,''~``readout circuit,''~``sense-node gain,''~``preceding-stage output gain'' and `` chip output gain'' are indicated.}
	\label{fig:xrpix2b_circuit}
	\vspace*{-0.5\intextsep}
\end{figure}

\begin{figure}[!tb]
	\centering
	\vspace*{-2\intextsep}
	\includegraphics[width=16cm]{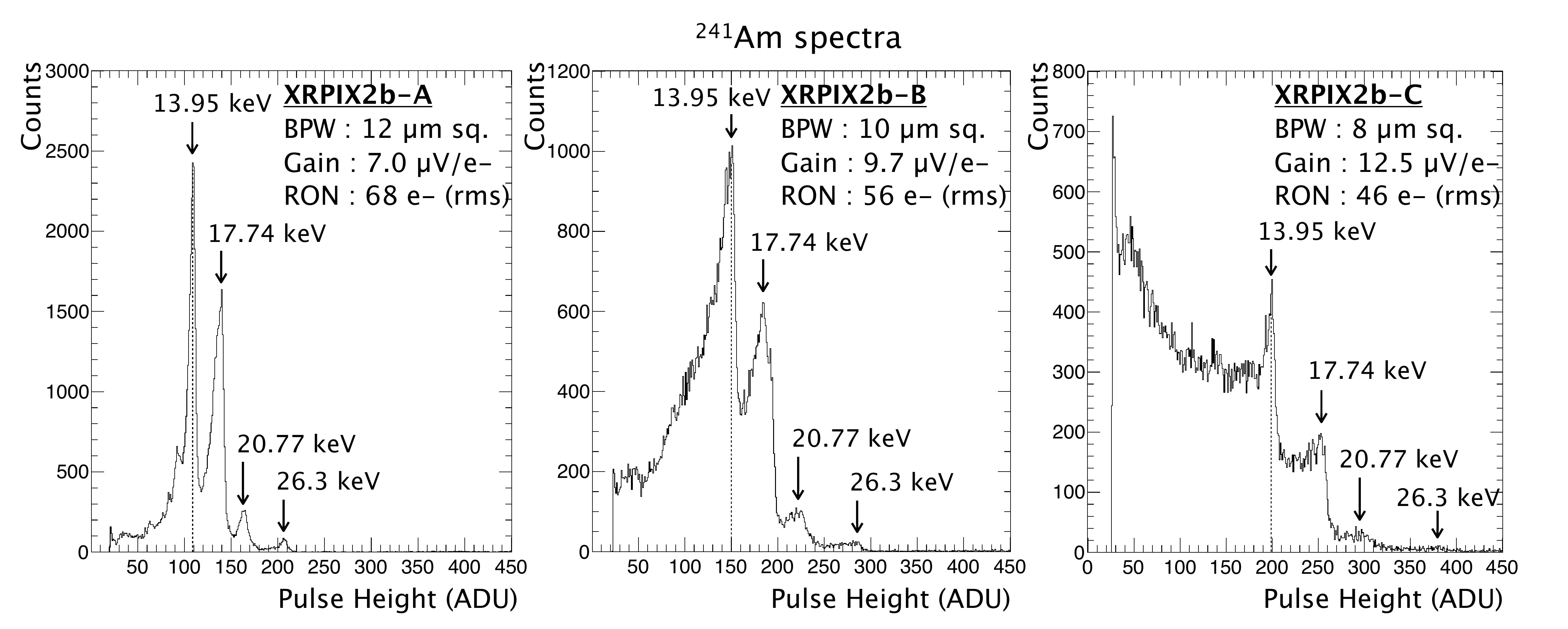}
	\vspace*{-1.5\intextsep}
	\caption{X-ray spectra of an ${\rm ^{241}}$Am radioisotope obtained with the two TEGs in XRPIX2b. The pulse height is shown in analog digital units (ADU). 1 ADU is 244 ${{\rm \mu V/}e^{-}}$ (1~V~/~12~bit).}
	\label{fig:xrpix2b_diff-spectra}
	\vspace*{1.5\intextsep}
\end{figure}% 

\begin{table}
	\begin{center}
	\vspace*{-\intextsep}
	\caption{Specification Summary of XRPIX2b.}
	\label{table:xrpix2b_list}
	\begin{tabular}{ccccc} \hline
		TEG Name & \# of Pixel & BPW Size (${\rm \mu m~sq.}$) & Gain (${{\rm \mu V/}e^{-}}$) & Readout Noise (${e^{-}~{\rm rms}}$) \\ \hline
		A & 144 ${\rm \times}$ 144 (${\rm \sim}$21k) & 12 & 7.0 & 68 \\
		B & 144 ${\rm \times}$ 8 (${\rm \sim}$1k) & 10 & 9.7 & 56 \\
		C & 8 ${\rm \times}$ 144 (${\rm \sim}$1k) & 8 & 12.5 & 46 \\
		D & 8 ${\rm \times}$ 8 & 6 & 16.0 & 36 \\ \hline
	\end{tabular}
	\end{center}
\end{table}% 

The chip output gain and the readout noise of XRPIX1, the first prototype of the XRPIX series, were 3.6 ${{\rm \mu V/}e^{-}}$ and 129 ${e^{-}}$ (rms) \cite{ryu_tns11}, where the chip output gain and the readout noise are defined as the conversion gain from signal charge to output voltage at the output of the chip (Figure~\ref{fig:xrpix2b_circuit}) and the width of pedestal peak, respectively.
We found the parasitic capacitance of the sense-node is dominated by the capacitive coupling between a transistor layer and a buried p-well (BPW)~\cite{ryu_tns13}\cite{nakashima_nima13}. 
The BPW is the p-type dopant region in the sensor layer under the BOX layer (see Figure~\ref{fig:soipix}), which is one of the key technology of SOIPIX and is introduced as an electrical shield to suppress the back-gate effect~\cite{arai_nima11}.
In XRPIX1b following XRPIX1, we decrease the area of the BPW to 14 ${\rm \mu m}$ square and obtained higher chip output gain of 6.2 ${{\rm \mu V/}e^{-}}$ and better readout noise of 70~${e^{-}}$ (rms)~\cite{takeda_tns13}. 

In order to further increase the gain, we developed the XRPIX2b, which is the forth prototype of the XRPIX series (Figure~\ref{fig:xrpix2b_chip}). 
The detailed description is found in~\cite{takeda_pos14} and its pixel circuit is shown in Figure~\ref{fig:xrpix2b_circuit}. 
XRPIX2b has four kinds of test element groups (TEGs) with the same pixel circuit layout but with different BPW sizes (Table~\ref{table:xrpix2b_list}). 
We obtained spectra of X-rays from an ${\rm ^{241}}$Am radioisotope with the four TEGs in the frame readout mode, in which we read out analog signals from all pixels serially.
After the data reduction and analyses, the details of which are found in~\cite{ryu_tns11}, we obtained chip output gains and readout noises given in Table~\ref{table:xrpix2b_list}. 
The examples of the spectra (TEG--A and --C) is shown in Figure~\ref{fig:xrpix2b_diff-spectra}. 

\begin{figure}[!tb]
	\vspace*{\intextsep}
	\begin{minipage}{0.5\hsize}
 		\begin{center}
			\includegraphics[height=7.5cm]{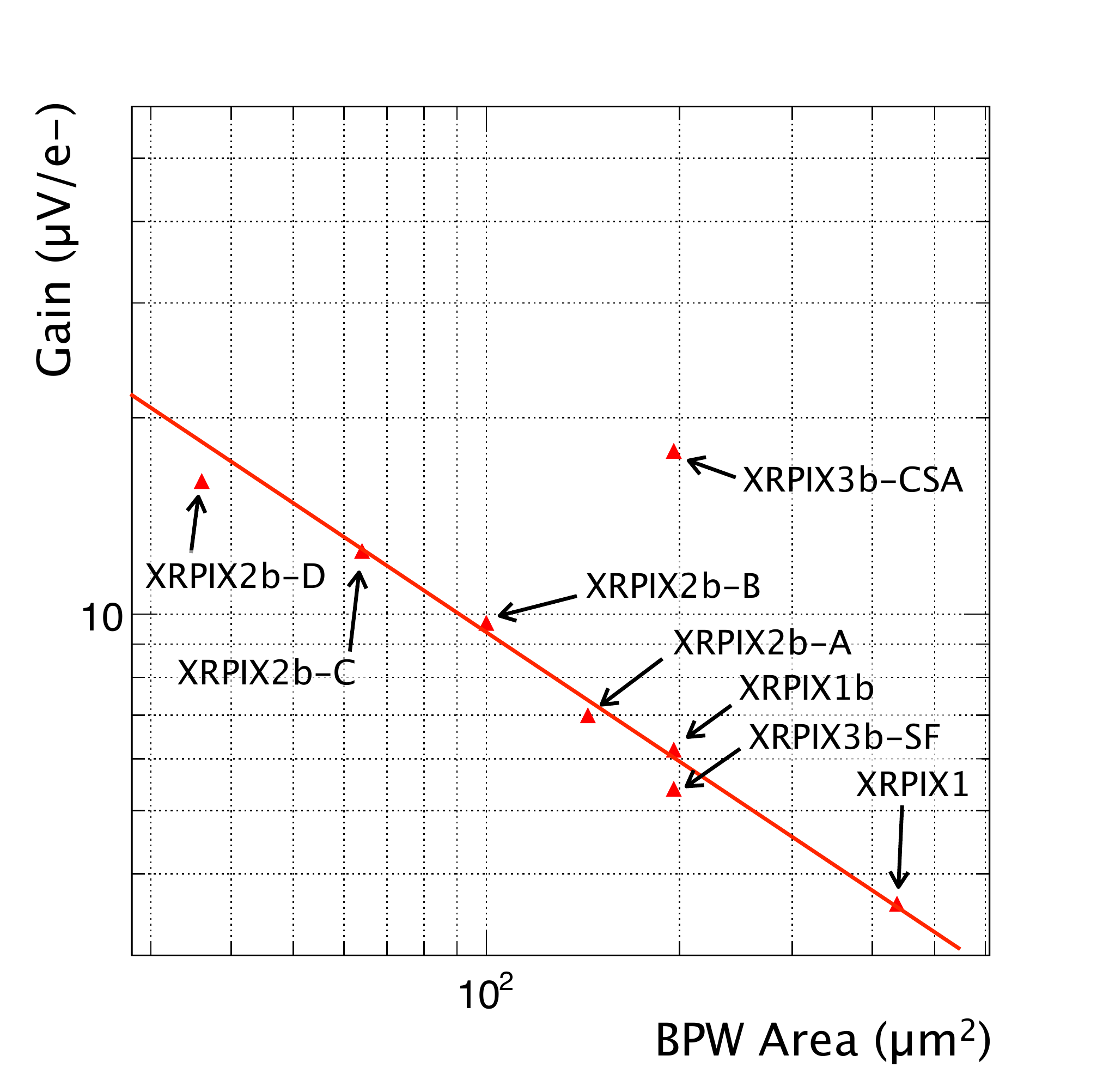}
			\end{center}
		\vspace*{-0.5\intextsep}
		\caption{Relation between the BPW area and the conversion gain.}
	 	\label{fig:bpw-area_gain}
	\end{minipage}
	\hspace{3mm}
	\begin{minipage}{0.5\hsize}
		\begin{center}
		\includegraphics[height=7.5cm]{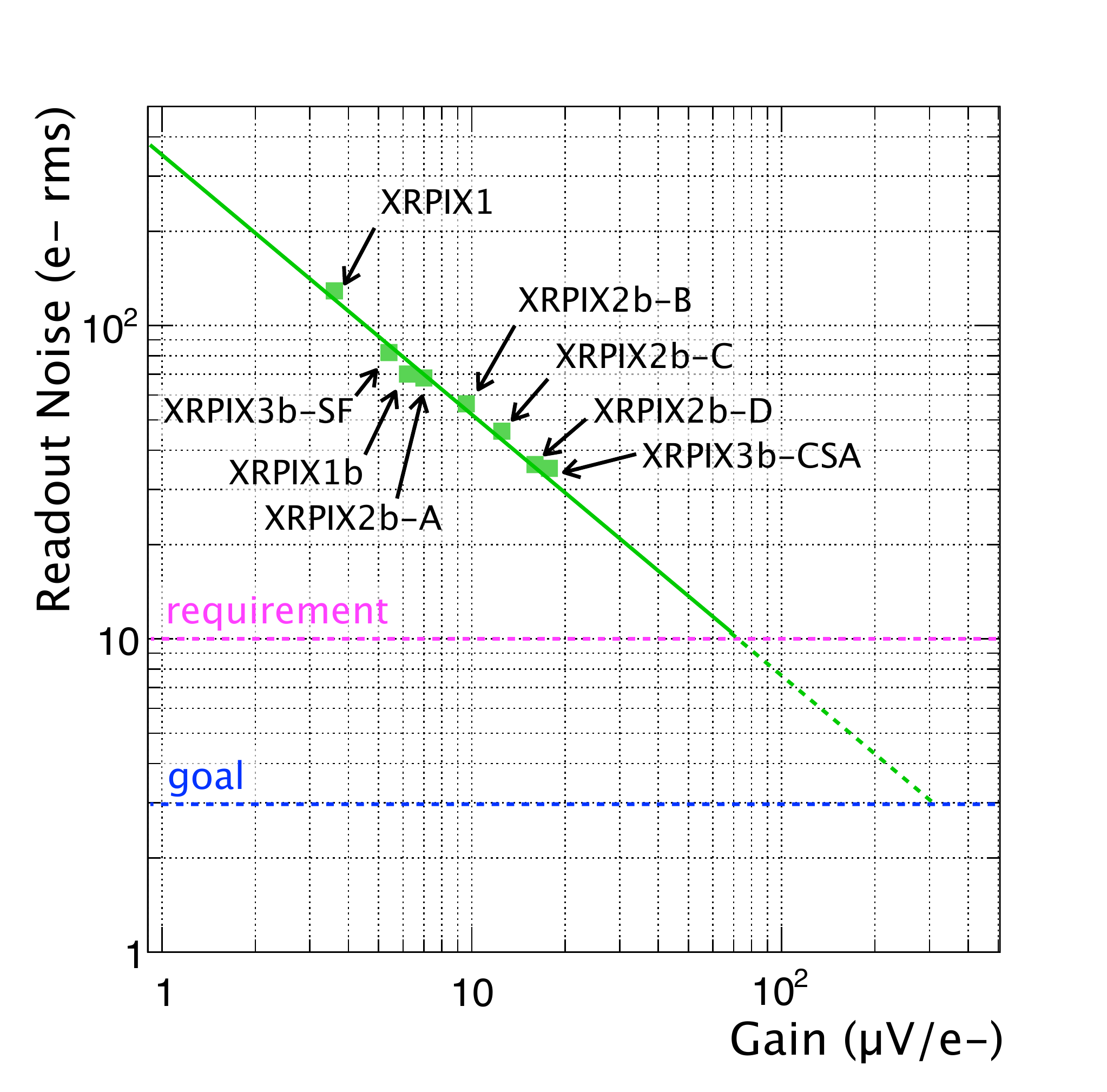}
		\end{center}
		\vspace*{-0.5\intextsep}
		\caption{Relation between the chip output gain and the readout noise.}
		\label{fig:gain_ron}
	\end{minipage}
	\vspace*{0.5\intextsep}
\end{figure}

We successfully increased the chip output gain and made further reduction of the readout noise by decreasing the area of the BPW.
Figure~\ref{fig:bpw-area_gain} shows the relationship between the BPW area and the gain, which is fitted with a power-law function.
We plot the readout noise as a function of the chip output gain for the TEGs in XRPIX2b, XRPIX1 and XRPIX1b in Figure~\ref{fig:gain_ron} 
and found the correlation can be fitted with a power-law function whose power is ${\rm \sim -}$1. 

Although we succeeded in the reduction of the readout noise by increasing the conversion gain as shown above, we encountered a new problem. 
As is shown in Figure~\ref{fig:xrpix2b_diff-spectra}, the X-ray spectral response of a single line has a significant tail structure especially in TEG--B and --C. 
The fraction of the tail component to the gaussian increases as the size of BPW decreases from TEG--A to --C. 
Through the previous studies on the tail structures by \cite{nakashima_nima13}\cite{matsumura_nima15}\cite{matsumura_nima14}, we consider that it is due to the loss of the signal charge in the sensor layer.
The BPW works also as a part of the sense node as well as an electrical shield since the BPW is electrically connected to the sense node. 
Thus, too small size of the BPW is unable to collects all the signal charge generated and degrades the charge collection efficiency (CCE) and spectral performance. 
This result suggests that there is a limit to the method of increasing the gain by reducing the area of the BPW in order to improve spectral performance.

\section{Improvement of Spectral Performance by CSA}
\label{sec:xrpix3b}

We introduced in-pixel CSAs in XRPIX3b, the sixth prototype of an XRPIX series, to raise the chip output gain by increasing the preceding-stage output gain keeping an appropriate area of the BPW.
In this section, we describe the device specification and spectral performance of XRPIX3b.

\subsection{Device Description}
\label{sec:device_description}

\begin{figure}[!tb]
	\vspace*{-\intextsep}
	\hspace{-2mm}
	\begin{minipage}{0.5\hsize}
 		\begin{center}
			\includegraphics[height=7.8cm]{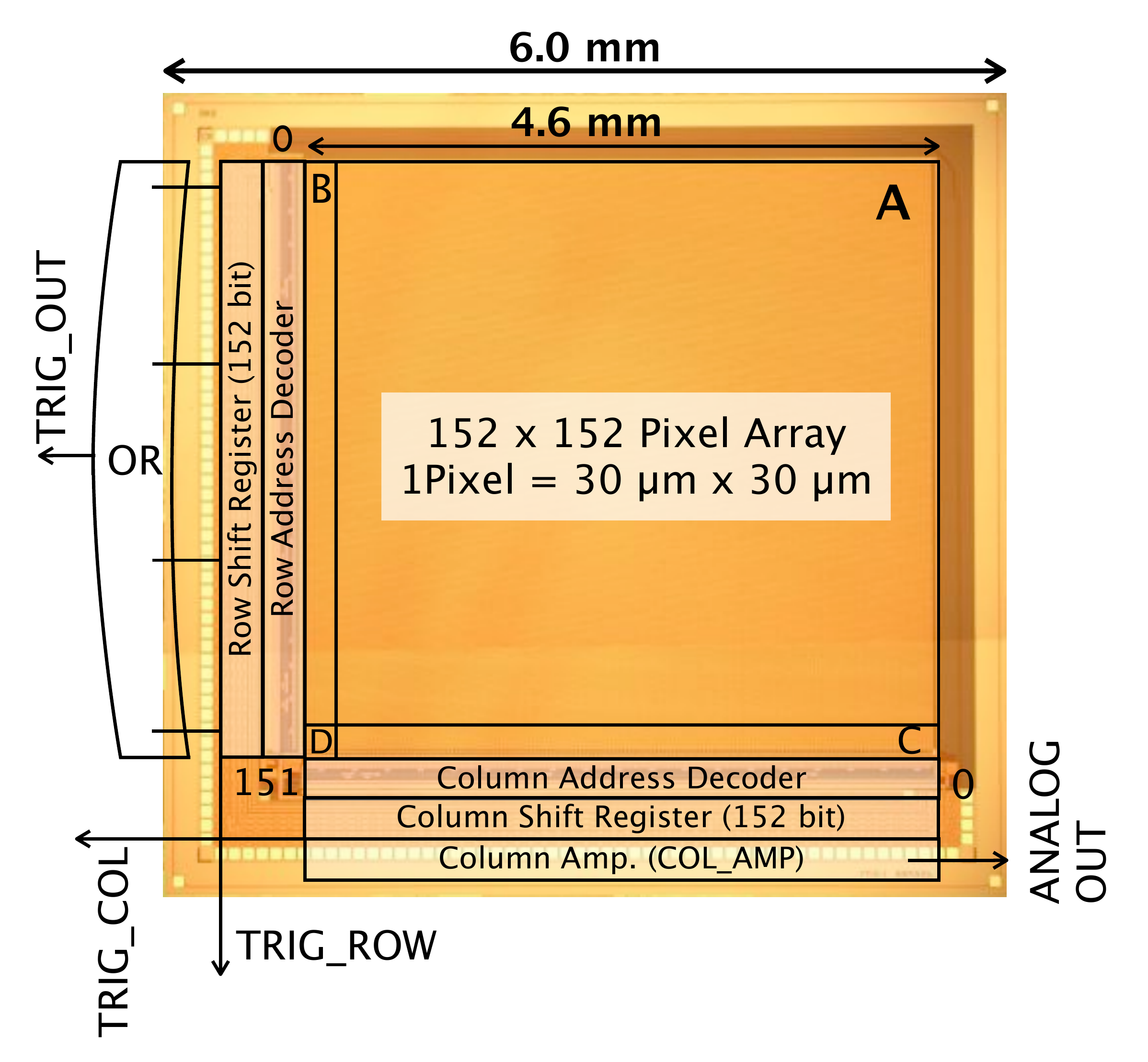}
		\end{center}
		\vspace*{-1\intextsep}
		\caption{Chip photograph of the XRPIX2b and its block diagram. The location of each TEG is indicated.}
	 	\label{fig:xrpix2b_chip}
	\end{minipage}
	\hspace{4mm}
	\begin{minipage}{0.5\hsize}
		\begin{center}
		\includegraphics[height=7.5cm]{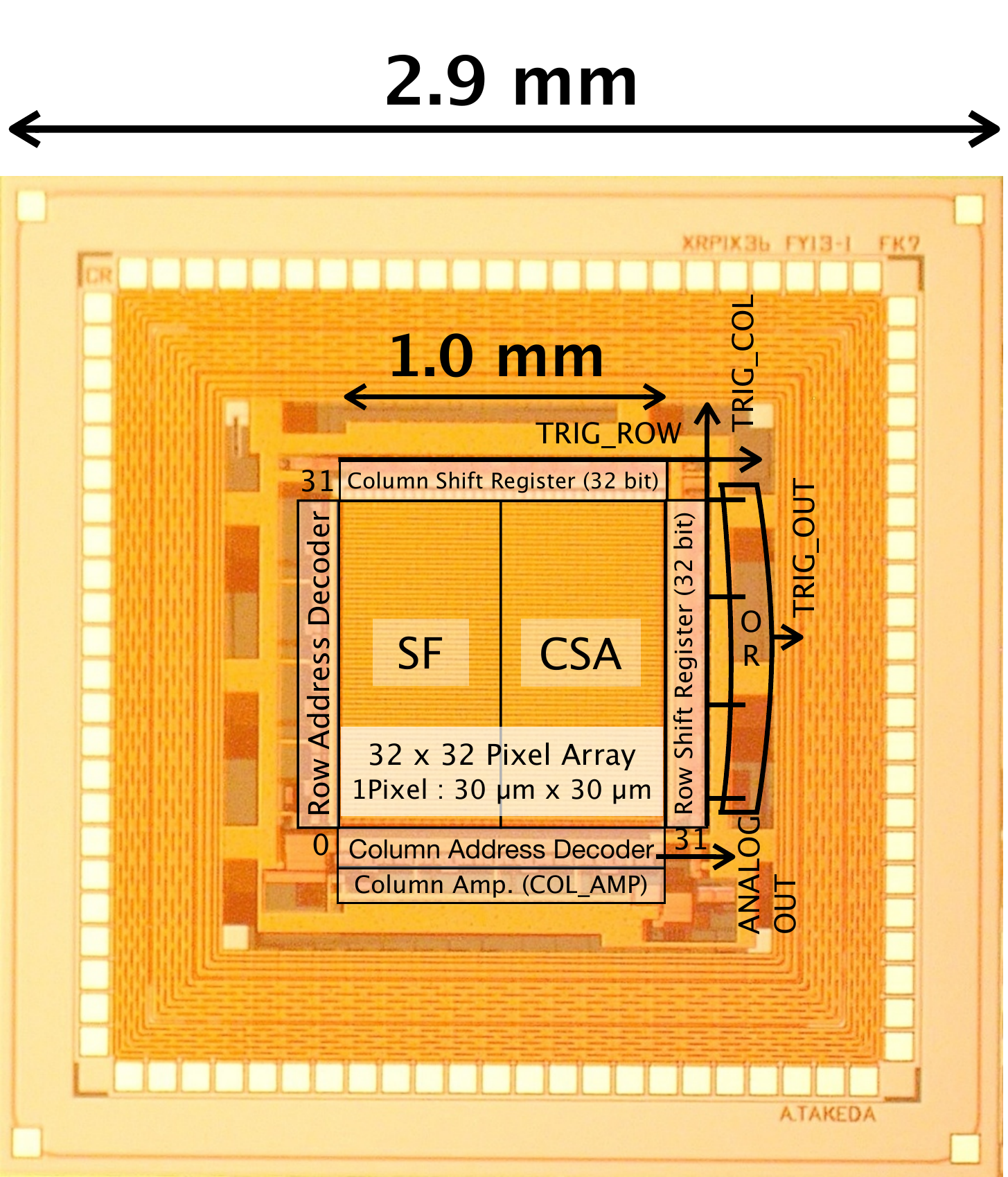}
		\end{center}
		\caption{Chip photograph of XRPIX3b and its block diagram. Two types of TEG are used.}
		\label{fig:xrpix3b_chip}
	\end{minipage}
\end{figure}

\begin{figure}[!tb]
	\centering
	\vspace*{\intextsep}
	\includegraphics[width=11cm]{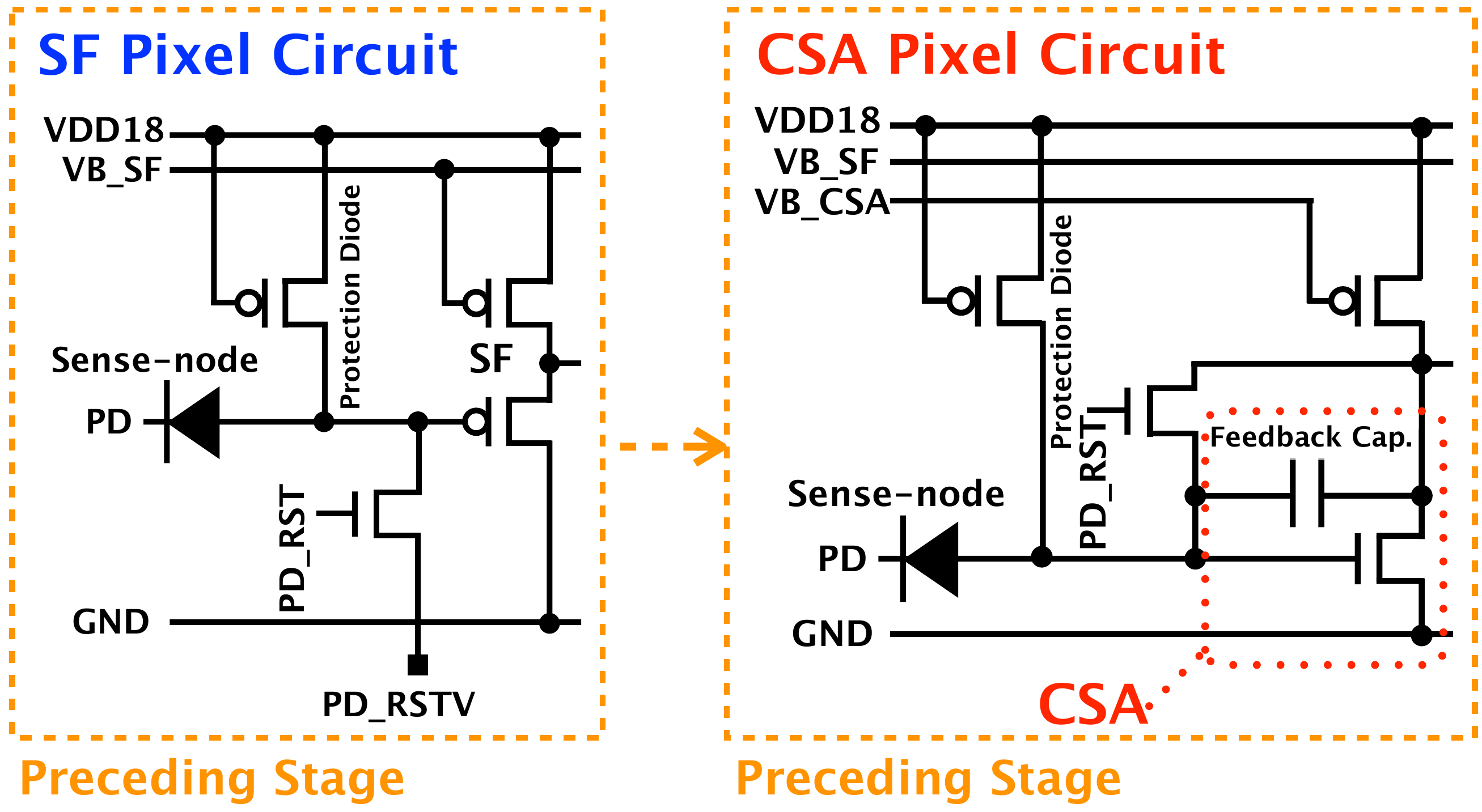}
	\caption{Pixel circuit of XRPIX3b. The basic component of the pixel circuit is same as the previous devices. XRPIX3b has CSA circuit newly. The feedback capacitor used in the CSA circuit is 1 fF.}
	\label{fig:xrpix3b_circuit}
	\vspace*{2\intextsep}
\end{figure}

XRPIX3b was fabricated using the 0.2 ${\rm \mu m}$ fully depleted SOI CMOS pixel process by Lapis Semiconductor Co., Ltd., the same process as XRPIX2b.
The detector is 2.9 mm ${\rm \times}$ 2.9 mm in size and consists of 32 ${\rm \times}$ 32 pixels.
The pixel size and the imaging area are 30 ${\rm \mu m~\times}$ 30 ${\rm \mu m}$ and approximately 1.0 mm ${\rm \times}$ 1.0 mm, respectively.
Based on the lesson learned, described in the Section~\ref{sec:spectroscopic_performance}, we adopted the BPW size of 14 ${\rm \mu m}$ in order to keep enough CCE.
Herein, we show a result of the n-type Czochralski-SOI wafer device with a thickness of 310 ${\rm \mu m}$. 

The sensor format and block diagram are shown in Figure~\ref{fig:xrpix3b_chip}.
The chip contains two different types of TEG pixels.
One of the two, ``CSA TEG,'' has in-pixel CSAs while the other, ``SF TEG'' has same pixel circuit but with SFs instead of CSAs, which are shown in the right and left sides of the device shown in Figure~\ref{fig:xrpix3b_chip}. 
Figure~\ref{fig:xrpix3b_circuit} shows the preceding-stage circuits used in the two TEGs where the feedback capacitance of 1~fF is used in the in-pixel CSA of the CSA TEG.

\subsection{Spectral Performance}
\label{sec:spectral_performance}

\begin{figure}[!tb]
	\vspace*{-2\intextsep}
	\begin{minipage}{0.5\hsize}
 		\begin{center}
		\vspace*{-\intextsep}
		\includegraphics[height=6.5cm]{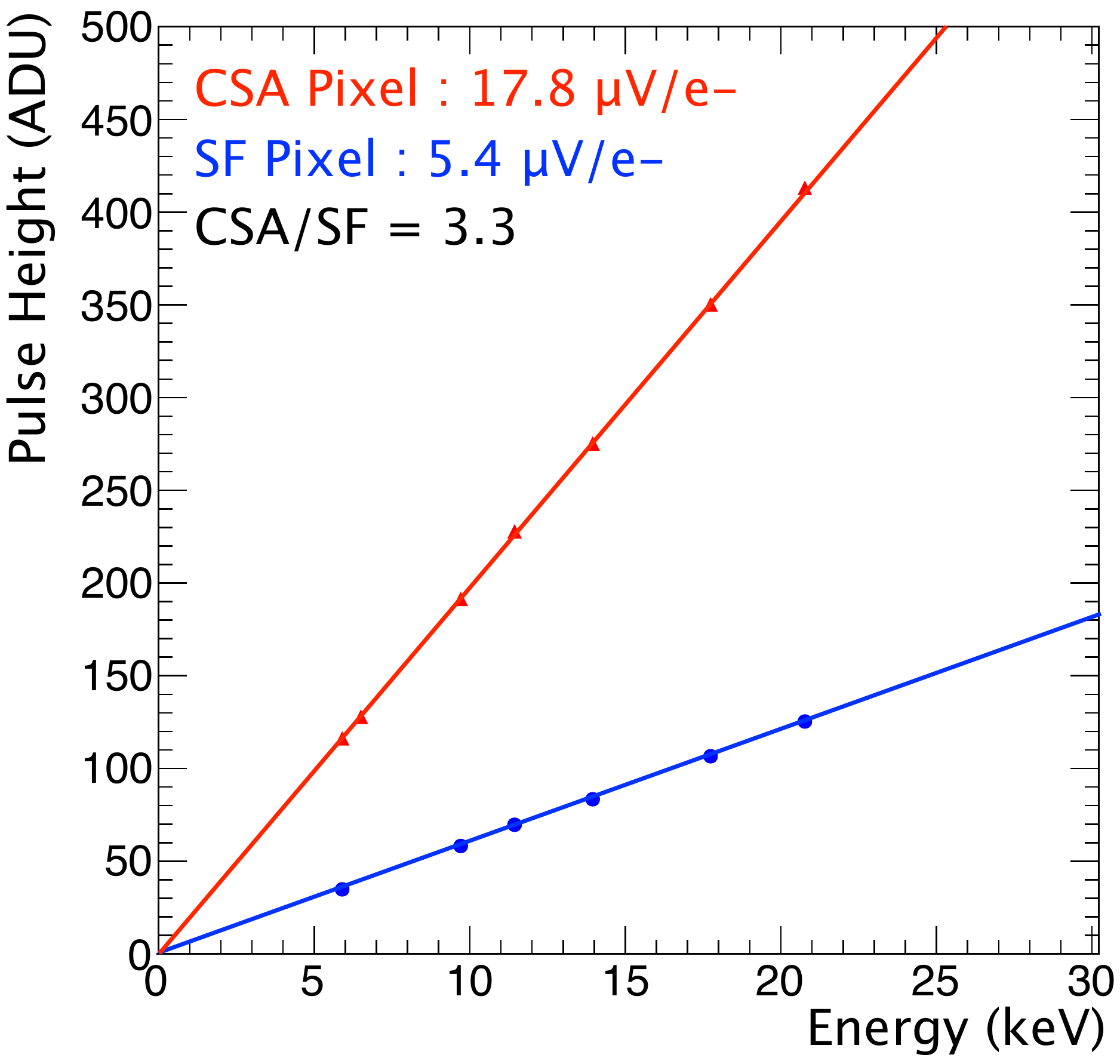}
		\end{center}
		\vspace*{-1.5\intextsep}
		\caption{Calibration plot of X-ray energy and signal pulse height.}
	 	\label{fig:xrpix3b_calib}
	\end{minipage}
	\hspace{2mm}
	\begin{minipage}{0.5\hsize}
		\begin{center}
		\includegraphics[height=6.5cm]{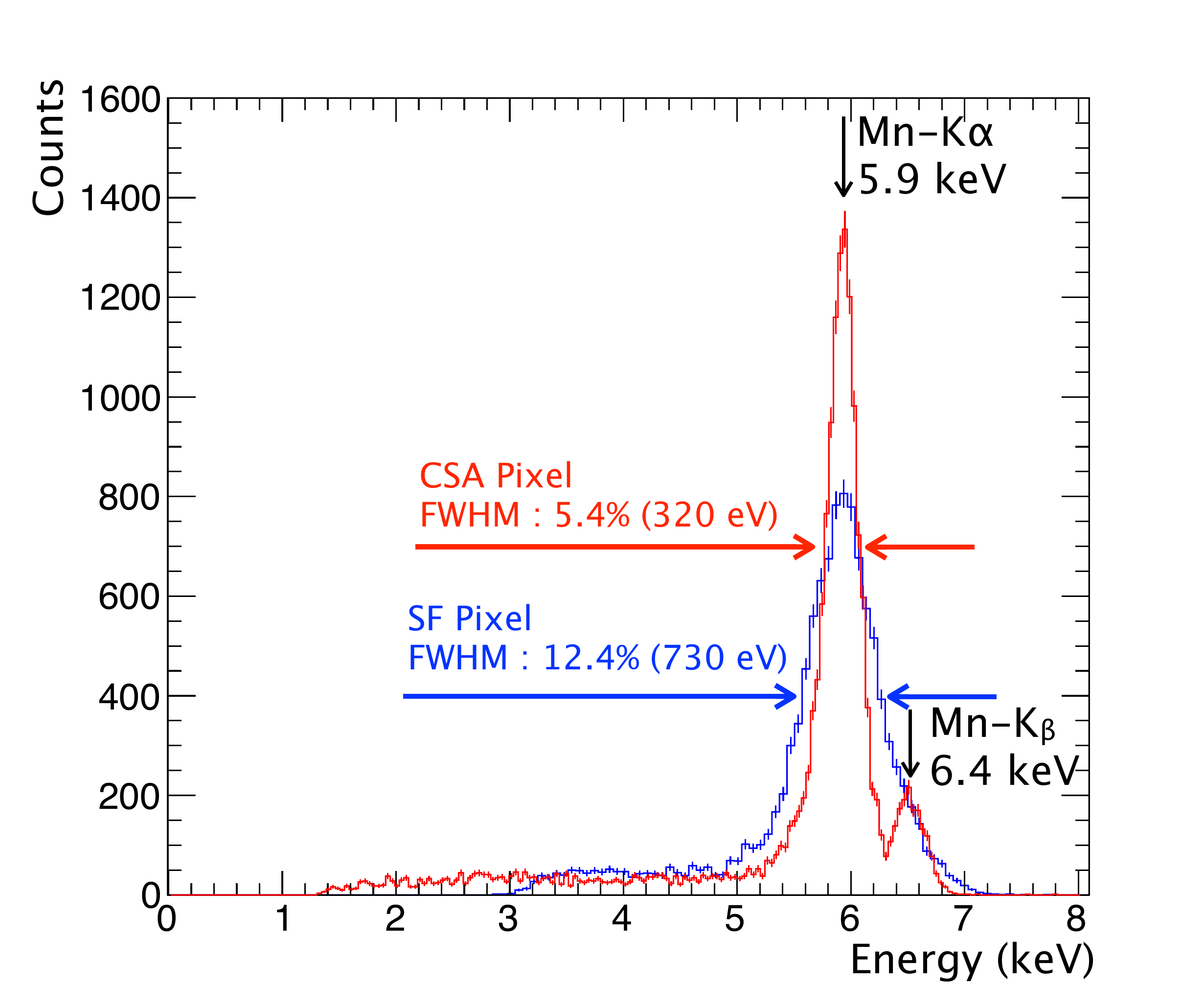}
		\end{center}
		\vspace*{-1.5\intextsep}
		\caption{Comparison spectra of the SF and CSA pixel circuits. The spectra are those of a ${\rm ^{55}Fe}$ radioisotope obtained in all pixel mode.}
		\label{fig:xrpix3_fe55}
	\end{minipage}
\end{figure}

Figure~\ref{fig:xrpix3b_calib} shows plots of X-ray energy calibrations of XRPIX3b 
using 5.9, 6.4, 9.71, 11.44, 13.95, 17.74, and 20.77 keV  X-ray lines from a ${\rm ^{55}}$Fe and an ${\rm ^{241}}$Am. 
The SF and CSA TEGs have the chip output gains of 5.4 ${{\rm \mu V/}e^{-}}$ and 17.8~${{\rm \mu V/}e^{-}}$, respectively. 
The CSA TEG has the chip output gain 3.3 times higher than the SF TEG. 

Figure~\ref{fig:xrpix3_fe55} shows the ${\rm ^{55}Fe}$ X-ray spectra of single pixel events obtained with the two TEGs after the data reduction and analyses given in \cite{ryu_tns11}.
No significant tail structures are seen in the spectra. 
The energy resolutions of the SF and CSA TEGs are 730 eV (12.4{\%}) and 320 eV (5.4{\%}) in FWHM at 5.9 keV, respectively. 
The readout noises of the SF and CSA TEGs are 82 and 35 ${e^{-}}$ (rms), respectively. 
We successfully improved the X-ray spectral performance by introducing the in-pixel CSA circuit without degradation of the CCE and resolved Mn-${\rm K{\alpha}}$ and -${\rm K{\beta}}$ lines for the first time in the XRPIX series.

\section{Discussion and Future Prospects}
\label{sec:future_prospects}

Plotting the readout noise and chip output gain of the CSA TEG of XRPIX3b in Figure~\ref{fig:gain_ron}, 
we found it follows the correlation between those of XRPIX1, 1b and the TEGs of XRPIX2b in spite that the preceding stage output gains are different. 
All the devices including XRPIX3b are equipped with the same readout circuits. 
This implies that almost all the readout noise is attributed not to the preceding stage but to the readout circuit. 
Thus, improving the readout circuit is significantly effective to reduce the readout noise. 
Increasing the preceding stage output gain is also effective. 

In this paper, we investigated the adoption of small BPW and in-pixel CSAs in order to improve the spectral performance of XRPIX.  
We found that the reduction of the size of the BPW surly increases the gain. 
However, there is a limit to it because too small BPW degrades the CCE. 
We consider the BPW size of ${\rm \sim}$14~${\rm \mu m}$ is optimal for the pixel size of ${\rm 30~\mu m}$. 
The in-pixel CSA also increases the preceding stage output gain and reduces the readout noise successfully. 
While the CSA with the feedback capacitance of 1 fF is expected to have the chip output gain of ${\rm \sim}$160~${{\rm \mu V/}e^{-}}$, the observed chip output gain is only ${\rm \sim}$${\rm 1/10}$ of that.
It suggests that there would be a room to optimize a parameter of the CSA.
In the following prototype devices, we will make further improvement of the spectral performance according to the guideline obtained in this paper.

\acknowledgments

We acknowledge the valuable advice and great work by the personnel of LAPIS Semiconductor Co., Ltd.
This study was supported by the Japan Society for the Promotion of Science (JSPS) KAKENHI Grant-in-Aid for Scientific Research on Innovative Areas 25109002 (Y.A), 25109003 (S.K) and 25109004 (T.G.T), Grant-in-Aid for Scientific Research (B) 23340047 (T.G.T) and Grant-in-Aid for Young Scientists (B) 25870347 (T.T).
This study was also supported by the VLSI Design and Education Center (VDEC), the University of Tokyo in collaboration with Cadence Design Systems, Inc., and Mentor Graphics, Inc.

\end{document}